# Quasi-ballistic heat conduction due to Lévy phonon flights in silicon nanowires

R. Anufriev[1,*], S. Gluchko[1,2], S. Volz[1,2], & M. Nomura[1,3]

[1] Institute of Industrial Science, the University of Tokyo, Tokyo 153–8505, Japan.
[2] Laboratory for Integrated Micro Mechatronic Systems / National Center for Scientific Research-Institute of Industrial Science (LIMMS/CNRS-IIS), the University of Tokyo, Tokyo 153–8505, Japan.
[3] PRESTO, Japan Science and Technology Agency, Saitama 332–0012, Japan.
[*] Correspondence should be addressed to R.A. (email: anufriev@iis.u-tokyo.ac.jp) or M.N. (email: nomura@iis.u-tokyo.ac.jp).

*Future of silicon-based microelectronics relies on solving the heat dissipation problem. A solution may lie in a nanoscale phenomenon known as ballistic heat conduction, which implies heat conduction without heating the conductor. But, attempts to demonstrate this phenomenon experimentally are controversial and scarce whereas its mechanism in confined nanostructures is yet to be fully understood. Here, we experimentally demonstrate quasi-ballistic heat conduction in silicon nanowires (NWs). We show that the ballisticity is strongest in short NWs at low temperatures but weakens as the NW length or temperature is increased. Yet, even at room temperature, quasi-ballistic heat conduction remains visible in short NWs. To better understand this phenomenon, we probe directionality and lengths of phonon flights. Our experiments and simulations show that the quasi-ballistic phonon transport in NWs is the Lévy walk with short flights between the NW boundaries and long ballistic leaps along the NW.*

## Introduction

We witness the final clash between Moore's law and its old enemy – heat[1,2]. The first battles were lost in the early 2000s when overheating limited the clock speed of microprocessors[1,3]. Today, the processors face a new limit as growing power density makes it impossible to further reduce the transistor size without overheating the processor. Thus, the future of silicon-based microelectronics may depend not on further improvements of the fabrication process, but on a deeper understanding of nanoscale heat conduction.

Fortunately, at nanoscale heat can be conducted without heating the conductor. This phenomenon, known as ballistic heat conduction, occurs when phonons ballistically travel between scattering events for hundreds of nanometres without energy dissipation[4]. The fully ballistic heat conduction implies that the thermal conductance ($K$) of a nanostructure is independent of the structure length ($L$), thus the thermal conductivity ($\kappa \propto K \times L$) loses its meaning as a material property and becomes merely proportional to the length. Such a length-dependent thermal conductivity was experimentally observed at room temperature in silicon films[5] and membranes[6], SiGe and $Ta_2Pb_3Se_8$ nanowires (NWs)[7–9], and suspended graphene[10,11]. However, Chang with colleagues argued[7,8,12] that the thermal conductivity may depend on the length due to the thermal contact resistance, even if heat conduction is purely diffusive. They reanalyzed[8,12] the data of the experimental works and concluded that heat conduction in silicon shows no apparent signs of ballisticity. Indeed, recent measurements[8,13–16] on silicon NWs seem to confirm that heat conduction at room temperature is diffusive, with ballistic contribution visible only at low temperatures[13]. Nevertheless, theoretical works[17–20] insist that the ballistic heat conduction should be possible in short silicon NWs even at room temperature because the phonon mean free path in bulk spans up to ten micrometres[21–24]. Thus, the ballistic heat conduction in confined silicon nanostructures remains an interesting fundamental issue and a crucial aspect of silicon-based microelectronics.

Here, we study heat conduction in silicon NWs of different lengths and shapes using the micro time-domain thermoreflectance (µTDTR). Our measurements show how quasi-ballistic heat conduction occurs at low temperatures and weakens as the temperature is increased yet remains present even at room temperature. Moreover, we experimentally measure directionality and lengths of phonon flights and perform Monte Carlo simulations to revieal the mechanism of ballistic heat conduction.

## Experimental technique

To study the thermal properties of NWs, we use an all-optical µTDTR technique developed for contactless measurements on suspended nanostructures. Figure 1a shows the schematic of our experimental setup. While the continuous-wave probe laser monitors changes in the reflectance coefficient of an aluminium transducer, pulses of the pump laser periodically heat up the transducer, thus changing its reflectance coefficient. The transducer is located on a silicon island, which is suspended on four NWs (Fig. 1a). As heat dissipates through the NWs, the transducer cools down to the initial temperature, and the intensity of the reflected probe beam gradually returns to the background level. This decay of the probe signal can always be fitted by $exp(-t/\tau)$, where $t$ is the time since the end of the pump pulse and $\tau$ is the thermal decay time – the characteristic time of heat dissipation through the NWs being measured (Supplementary Fig. 1).

The measured thermal decay time is not directly linked to the thermal conductivity of the measured NWs because of the the thermal resistance of the silicon island, the aluminium transducer, and the heat sink. To eliminate the impact of the parasitic resistances and extract the thermal conductivity of the NWs, we simulate the same experiment for each sample using the Finite Element Method (FEM), implemented by Comsol Multiphysics. In the simulation, we use the thermal conductivity of NWs as a free parameter to find a value of thermal conductivity at which the simulated and measured decay curves match precisely, thus extracting the thermal conductivity of the measured NWs (Supplementary Fig. 1).

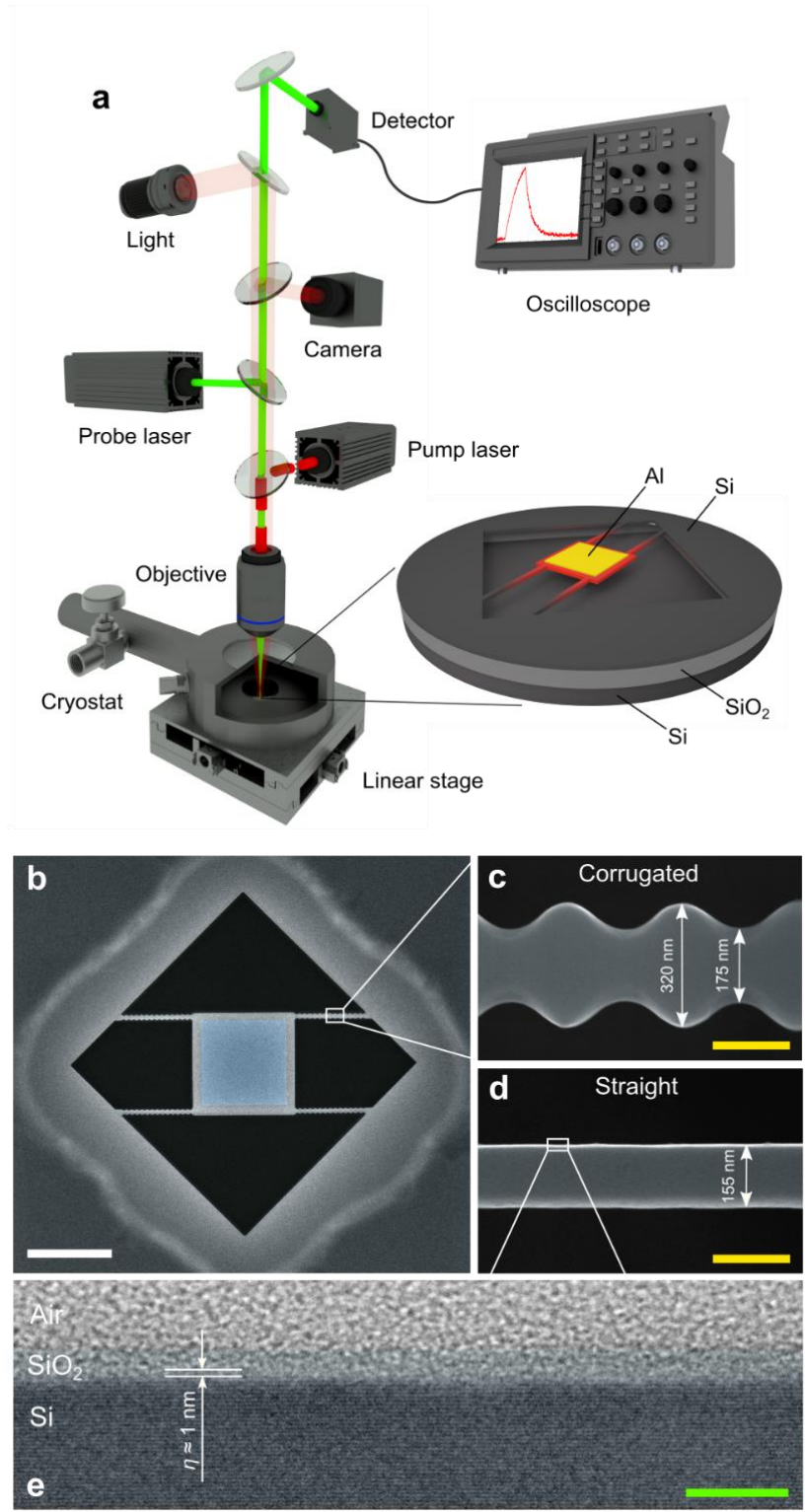

Figure 1. (a) Schematics of our μTDTR setup and a typical sample. SEM images show (b) a typical sample and close up views of (c) corrugated and (d) straight NWs. (e) TEM image of a NW boundary shows that the surface roughness ($\eta$) is a few atomic layers. Scale bars are (b) 5 μm, (c, d) 200 nm, and (e) 5 nm.

The samples were simultaneously fabricated on a silicon-on-insulator wafer using conventional top-down fabrication techniques (Methods). Figure 1b shows the scanning electron microscope (SEM) image of one of the samples. The NWs were of two types: corrugated (Fig. 1b), with the width of the narrow region of 175 nm and the width of the wide part of 320 nm, and straight (Fig. 1c), with the width of 155 nm. The thickness of all NWs was 145 nm. The transmission electron microscopy (TEM) image (Fig. 1e) shows that the surface roughness of NW side walls is only a few atomic layers (Supplementary Fig. 2). Since the TEM sample was inevitably exposed to air for a long time, a 2 nm oxide layer formed on the NW surface. However, the µTDTR samples were treated to minimise the air exposure and probably had a thinner oxide layer.

To reduce the experimental uncertainty, we fabricated three copies of each sample. Thus, each data point in this work is an average of the measurements on three different samples, with the error bars showing the standard deviation. For the error bars in thermal conductivity plots, we also add 1% due to by the uncertainty of the SEM measurements of NW dimensions.

**Length dependence of thermal properties**

First, we study how the thermal conductivity of straight and corrugated NWs depends on their length ($L$) and temperature. At the temperature of 4 K, the thermal conductivity decreases as the NWs become shorter for both straight and corrugated NWs (Fig. 2a). Such length-dependent thermal conductivity is the first sign of ballistic heat conduction effects. The slope of the length dependence for NWs of both types is proportional to $L^{0.3}$. As the temperature is increased to 100 K (Fig. 2b), the slope becomes proportional to $L^{0.25}$, yet the trends remain visible. At 200 K (Fig. 2c), the slope for the straight NWs further reduces to $L^{0.15}$, whereas the slope for corrugated NWs flattens. Finally, at room temperature (Fig. 2d), the length dependence for corrugated NWs becomes almost flat, whereas an $L^{0.13}$ trend is still visible for the straight NWs. An additional set of samples confirms this result (Supplementary

Fig. 3). In the absolute values, the corrugated NWs have 17% lower thermal conductivity than the straight NWs at 4 K. But, at 100K, the difference reduces to only 10%, then to 8 % at 200 K, and finally to about 5% at room temperature.

However, the length-dependence of thermal conductivity can be caused by the thermal contact resistance[8,12]. Thus, we should also examine the length dependence of the thermal resistance per unit area, given by $A / K$, where $A$ is the cross-section area and $K$ is the thermal conductance. Figure 3 shows that the data points for the long NWs ($L > 2.5$ μm) form a linear trend for both types of NWs and at all temperatures, which implies that heat conduction in the long NWs is diffusive[8,15,16]. This linear trend extrapolates precisely to zero resistance at zero length (Supplementary Fig. 4), showing that the contact resistance is negligible[8,12]. However, the data points for the short NWs ($L < 2.5$ μm) deviate from this linear trend. Such nonlinearity indicates the presence of ballistic heat conduction, which is the strongest at 4 K but weakens as the temperature is increased. Nevertheless, some degree of ballistic heat conduction is present even at room temperature.

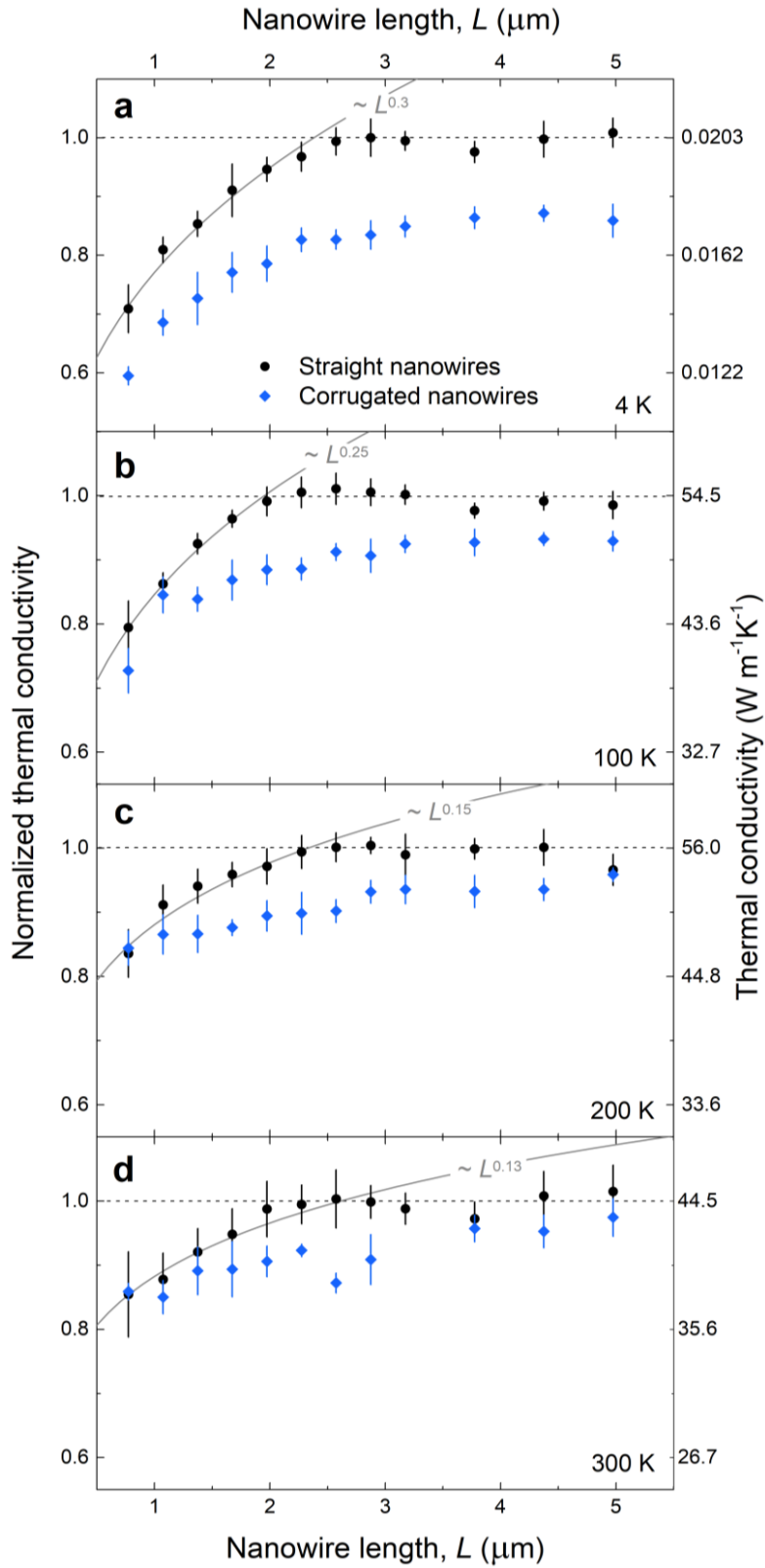

Figure 2. Length dependence of thermal conductivity. Whereas the thermal conductivity of long NWs is constant, the thermal conductivity of short NWs depends on the NW length. This dependence is different at the temperatures of 4 K (a), 100 K (b), 200 K (c), and 300 K (d). For convenient comparison, we normalised the thermal conductivity by the value at the plateau for straight NWs at each temperature. The error bars indicate the standard deviation in the measurements on three identical samples plus 1% uncertainty caused by the inaccuracy of SEM measurements.

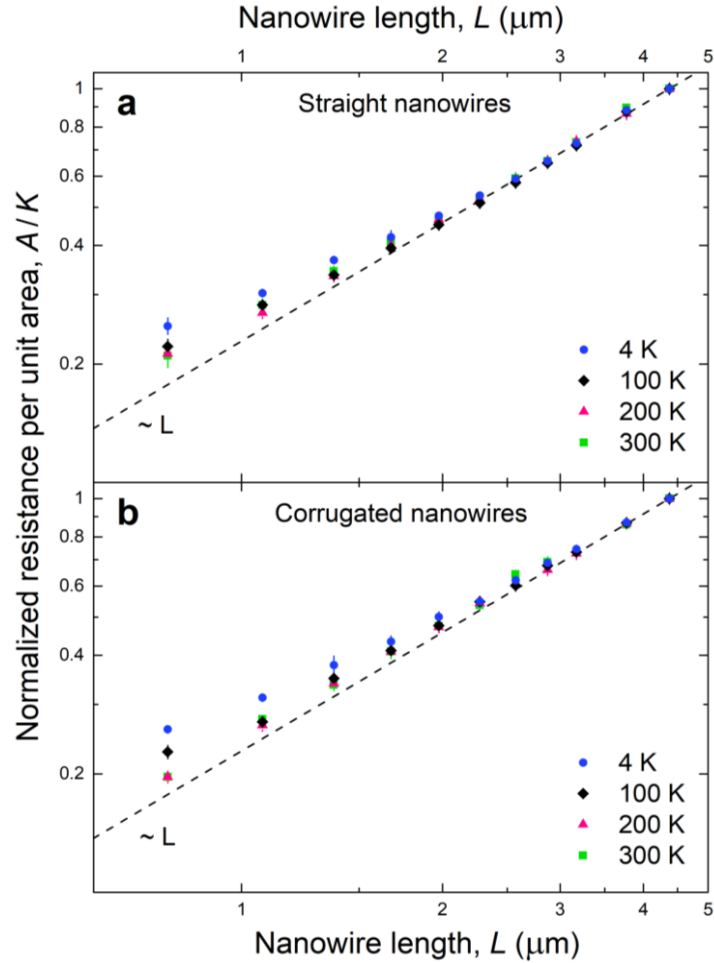

Figure 3. Nonlinear length dependence of thermal resistance. The thermal resistance per unit area, normalised by the value for the longest NWs at each temperature, shows nonlinear dependence on the length for both (a) straight and (b) corrugated NWs. This nonlinearity reveals quasi-ballistic heat conduction and is evident at 4 K but weakens as the temperature is increased, yet remains visible even at room temperature. The error bars indicate the standard deviation in the measurements on three identical samples plus 1% uncertainty caused by the inaccuracy of SEM measurements.

## Straight and serpentine nanowires

Another way to demonstrate quasi-ballistic heat conduction is to compare the NWs of the same length, width, and volume, but of two different shapes: straight (Fig. 4a) and serpentine (Fig. 4b). If phonons travel ballistically for a few micrometres in the straight NWs, then their ballistic path in the serpentine NWs should be limited by the length of one segment (700 nm). Thus, the thermal conductivity of the serpentine NWs should be lower. Indeed, at the temperature of 4 K, we measured a 32% lower thermal conductivity in the serpentine NWs (Fig. 4c). This result is consistent with the data measured by Heron *et al.*[25] in a similar experiment on serpentine NWs with the segment length of 600 nm at low temperatures (Supplementary Note 1). However, as we increased temperature, the difference gradually weakened. Above 200 K, the thermal conductivity of the straight and serpentine NWs was the same ($\kappa_{\text{straight}} / \kappa_{\text{serpentine}} = 1$) within the experimental uncertainty.

This transition is even more fascinating in terms of the thermal decay time (Fig. 4d). At low temperatures, thermal decay through the straight NWs is faster due to the presence of ballistic heat conduction, which is blocked in the serpentine NWs, as illustrated in the inset in Fig. 4d. But, as heat conduction becomes mostly diffusive at higher temperatures, the thermal decay through the serpentine NWs becomes faster. This inversion occurs because in the diffusive regime heat flux follows the shortest path through the serpentine NW and reach the heat sink more quickly. The simulation under the Fourier law approximation using the FEM illustrates that the heat flux cuts corners, as shown in the inset of Fig. 4d and Supplementary Fig. 5. Similar corner cutting behaviour has recently been also observed for electrons[26].

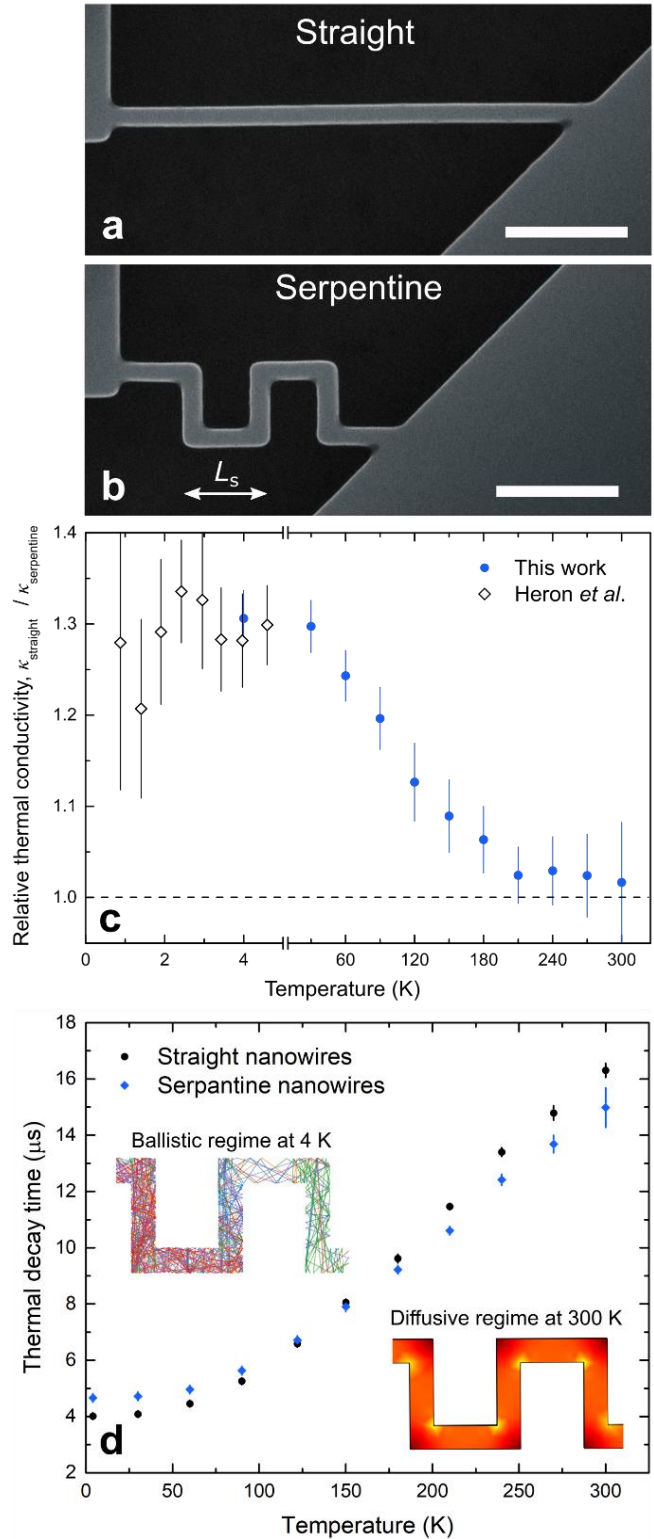

Figure 4. Heat conduction in straight and serpentine NWs. SEM images show (a) straight and (b) serpentine NWs. The length of the segment of the serpentine NW is 700 nm; the total length of both types of NWs is 4 μm. The scale bars are 1 μm. (c) The relative thermal conductivity ($\kappa_{straight} / \kappa_{serpentine}$) shows that the straight NWs have a higher thermal conductivity at low temperature, but the difference gradually disappears at higher temperatures. The error bars indicate the standard deviation in the measurements on three identical samples plus 1% uncertainty caused by the inaccuracy of SEM measurements. (d) The thermal decay time of straight and serpentine NWs at different temperatures show the transition from quasi-ballistic to the diffusive regime. The inserts show Monte-Carlo simulation in the ballistic regime and FEM simulation in the diffusive regime, in which heat fluxes cut corners.

**Directionality of phonon transport**

The observed lower thermal conductivity in the serpentine NWs implies that some phonons travel through the straight NWs experiencing very few diffuse scattering events. This is likely to occur if phonons fly through a NW along the most direct path from hot to cold. In our previous work[27], we demonstrated that narrow passages and periodic constrictions could form directional fluxes of phonons, which can ballistically travel for hundreds of nanometres in a given direction. Recent simulations support this possibility even at room temperature[28–30]. Such directional fluxes in our NWs, especially in those of the corrugated type, could explain the presence of ballistic heat conduction up to the room temperature.

To probe the directions of phonon transport, we developed the following experiment. The NWs were connected to the heat sink via a triangular detector consisting of a direct passage and a side arm, as shown in Fig. 5. The detectors were of two types: with direct (Fig. 5a) and indirect (Fig. 5b) connections to the heat sink. If phonons could form directional heat fluxes parallel to the NW axis, they would proceed to the direct passage rather than turned into the side arm. Then, in case of the direct detector, they could directly escape to the heat sink, whereas in case of the indirect detector they would have to take the longer path, as indicated by the arrows. Thus, comparing the heat dissipation time in samples with the direct and indirect detectors, we can probe the directionality of phonon transport. For this experiment, we fabricated a set of samples with the same detectors but different lengths ($L$), which allowed us to probe directionality of the phonon transport at different points of NWs. Since the directionality effects are stronger at low temperatures[27,28], and the low-temperature measurements are more stable and accurate, we conducted the experiments at 4 K.

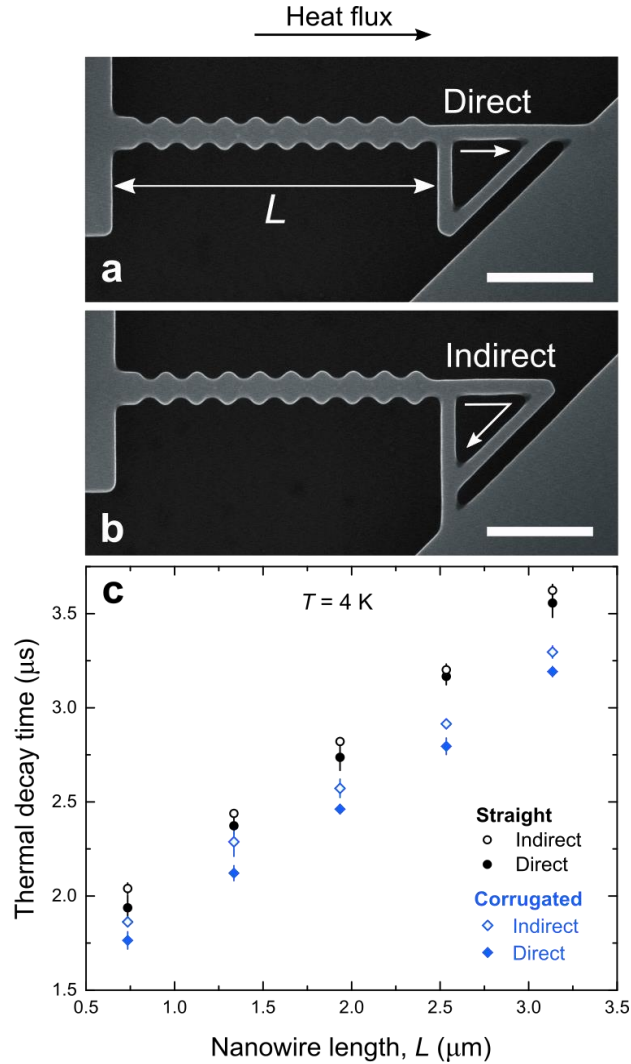

Figure 5. Directionality of phonon transport. SEM images show samples with the (a) direct and (b) indirect detectors. The scale bars are 1 µm. (c) The thermal decay time ($\tau$) measured on the samples with direct and indirect detectors is nearly the same in case of the straight NWs and only 5 – 10% different in case of the corrugated NWs. This result shows that the directions of phonon transport are almost isotropic. The error bars indicate the standard deviation in the measurements on three identical samples.

Figure 5c shows the measured thermal decay times. In straight NWs, the decay times ($\tau$) for the direct and indirect detectors are the same (within the experimental uncertainty) for all the lengths. Thus, the phonon transport directions in straight NWs are somewhat isotropic. However, in the corrugated NWs, the thermal decay is 5 – 10 % faster through the direct detector. An additional set of samples confirms this result (Supplementary Fig. 6). Thus, more phonons passed forward in the detector than turned into the side arm, which implies that phonons tend to be slightly oriented along the NW axis, as will be explained in the next section.

## Phonon transport simulations

To better understand our experimental results, we performed the Monte Carlo simulations of phonon transport (Methods). The algorithm records and analyses paths of phonons as they travel through the NWs (Supplementary Fig. 7). Once thousands of phonons are processed, we can obtain a statistical insight into the collective gas-like motion of phonons in our NWs. The simulations are conducted for the temperature of 4 K.

First, to address the apparent lack of directional phonon transport, we probe the angles at which phonons exit the NWs at the cold side. Figures 6a and 6b show distributions of the exit angles ($\theta$) for straight and corrugated NWs. In the straight NWs, the distributions have a dip at zero degrees due to the Lambert cosine distribution of the directions after diffuse scattering events at the NW side walls. The exit angle distribution for short NWs ($L = 0.8$ μm) is rather broad, but as the NWs become longer, the distribution slightly narrows (Fig. 6a). The narrowing occurs because the diffuse scattering at the surfaces randomises the phonon directions, whereas the structure geometry favours the phonons with the angle $\theta$ closer to zero degrees, as they are more likely to reach the end of the NW without another surface scattering. Such a process of randomisation and selection affects all the phonons in the system so that initial phonon gas evolves into a directional flux. However, even in the long NWs, the angular distribution remains rather broad, which is why we could not observe any clear directionality of phonon transport in straight NWs.

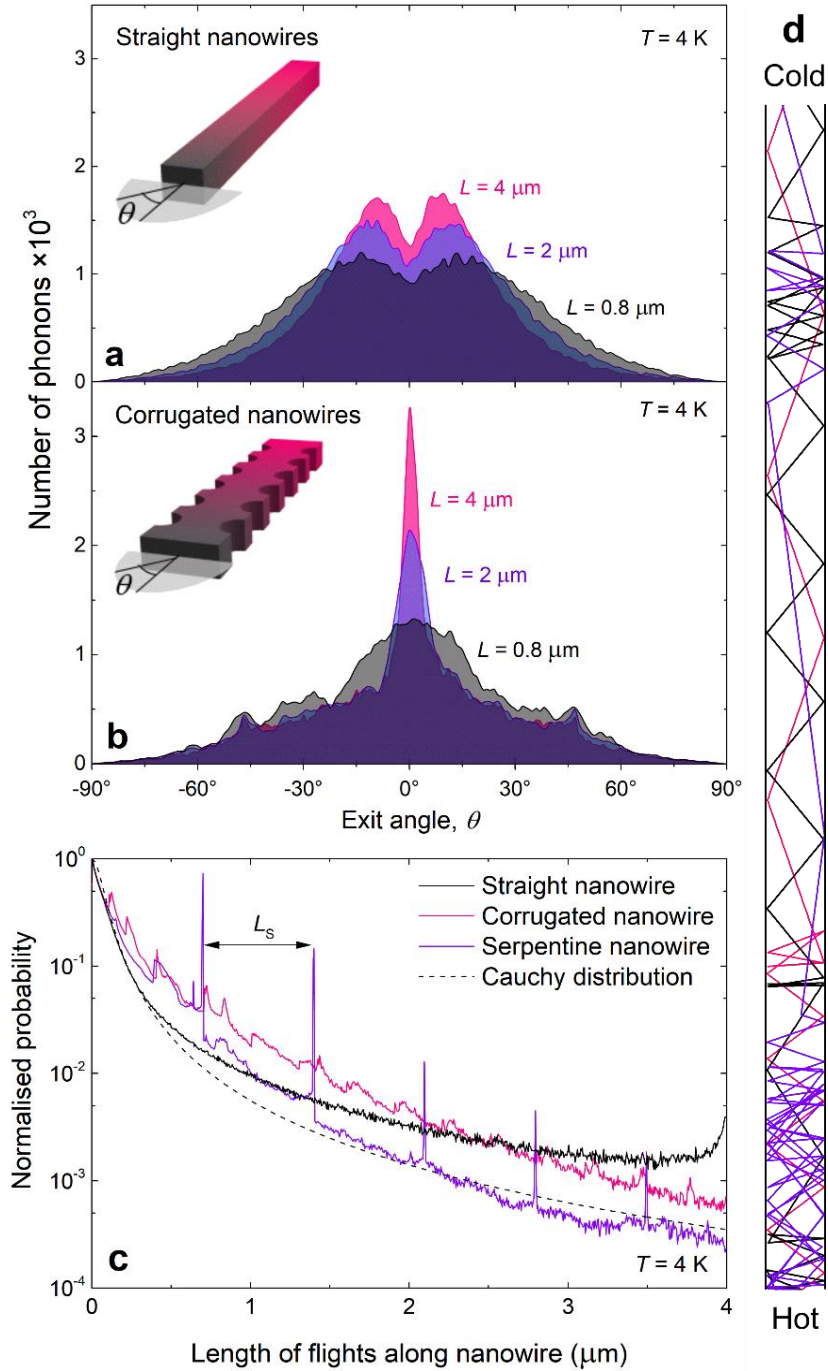

Figure 6. Monte Carlo simulation of phonon transport. Distribution of the angles at which phonons exit from the (a) straight and (b) corrugated NWs becomes narrower as NWs become longer. (c) The normalised probability of the length of ballistic flights along the NW axis in 4-µm-long NWs show the heavy-tailed distributions. (d) Typical simulated phonon trajectories in a straight NW illustrate the Lévy walk behaviour.

In the corrugated NWs (Fig. 6b), the process of randomisation and selection produces much narrower peaks of the exit angle distribution. This narrowing happens due to the very shape of corrugated NWs, which applies a stronger selection pressure on the phonon directions. Indeed, as the corrugations are likely to scatter the phonons backwards, only the phonons with directions that allow passing between the corrugated walls are likely to leave the NW. Thus, as the number of corrugations increases, the selection pressure strengthens and the exit angle distribution narrows, as shown in Fig. 6b. However, even the narrowest peak contains less than a quarter of all phonons, whereas most phonons still have a wide distribution of angles (Supplementary Fig. 8). For this reason, our experiments show only slight directionality of phonon transport in the corrugated NWs.

Next, we analyse how far phonons can ballistically fly in our NWs. Figure 6c compares probabilities of the ballistic phonon flights along the NW axis in 4-µm-long NWs of different types. All distributions have a maximum near zero, showing that phonons often bounce back and forth in the directions perpendicular to nanowire axis. The curve for the straight NW is a heavy-tailed distribution which resembles Cauchy distribution. Such a distribution is characteristic of the Lévy walk process, in which bursts of short flights are followed by long leaps, as illustrated in Fig. 6d. Remarkably, in the straight NW, the small increase around 4 µm shows that some phonons can ballistically traverse the entire NW. Conversely, in the serpentine NW, these long flights are blocked by the NW turns. Hence the peaks at the multiples of the segment length ($L_s$ = 700 nm). Moreover, the turns scatter phonon back into previous segments, which increases the likelihood of short flights (< 700 nm). Thus, fewer phonons can ballistically travel for a few micrometres, and the tail of the distribution is almost one order of magnitude lower.

In the corrugated NW, the situation is more complicated. On the one hand, phonons often become trapped within the corrugated boundaries, which multiplies the short flights. On the other hand, the scattering by the corrugated boundaries changes phonon directions and selects those parallel to the NW axis, as discussed above (Fig. 6b). Phonons that passed this selection can move forward between the corrugations for long distances. Thus, the distribution also has a heavy long-length tail like in the straight NW.

**Discussion**

In this work, we studied the ballistic heat conduction in silicon NWs at various length scales and temperatures. We demonstrated that at 4 K the thermal conductivity of both straight and corrugated NWs has an $L^{0.33}$ length dependence, the same as observed by Maire et al.[13] on straight NWs. The $L^{0.33}$ trend means that the heat conduction is neither fully ballistic ($L^1$) nor diffusive ($L^0$) but quasi-ballistic, meaning that phonons can ballistically traverse a part of a NW, but experience some diffuse scattering events as well. The fully ballistic heat conduction ($L^1$) can probably occur only in very short NWs or at sub-Kelvin temperatures, at which phonons only reflect at the surfaces elastically. However, the quasi-ballistic heat conduction is not uniquely a low-temperature phenomenon. Remarkably, even at room temperature, we measured $\kappa \propto L^{0.13}$ dependence, which agrees with the predicted $L^{0.1}$ for 1-µm-long silicon NWs[18]. For the shortest NW ($L = 775$ nm), we can estimate the ballistic thermal conductance as $G_{bal} = (L / \kappa_{total} - L / \kappa_{diff})^{-1}$. Taking the plateau value as $\kappa_{diff} = 45$ Wm$^{-1}$K$^{-1}$ and measured $\kappa_{total} = 38$ Wm$^{-1}$K$^{-1}$, we obtain $G_{bal} = 3.17 \times 10^{-8}$ Wm$^{-2}$K$^{-1}$, which is consistent with the predicted[18] value of $2.0 \times 10^{-8}$ Wm$^{-2}$K$^{-1}$.

As suggested by Chang with colleagues[7,8,12], we analysed the length dependence of the thermal resistance per unit area. This analysis revealed a negligible contact resistance and non-linear length dependence of the NW thermal resistance. These results further strengthen our conclusion about quasi-ballistic heat conduction in silicon in the 4 – 300 K range.

To demonstrate quasi-ballistic heat conduction even more directly, we compared straight and serpentine NWs. At 4 K, we found that the serpentine NWs are 32% less conductive, which is consistent with ~30% measured by Heron et al.[25] in a similar experiment. However, this difference gradually weakened at higher temperatures and disappeared at 200 K. Thus, at 200 K phonons already experience so many diffuse scattering events during their journey through the 700-nm-long segments that the additional scattering at the turns is insignificant. Nevertheless, recent simulations[18] show that if the segments were shorter (225 nm), a 14% difference between straight and serpentine NWs would still be visible even at room temperature, because many phonons travel ballistically for a few hundred nanometers.

Comparing straight and corrugated NWs, we found that the thermal conductivity of corrugated NWs is 17% lower at 4 K, which is consistent with measurements by Blanc *et al.*[31] on similarly corrugated NWs at the same temperature. The difference in the thermal conductivity of the straight and corrugated NWs correlated with ballisticity of heat conduction. As we increased the temperature, the difference became weaker. At room temperature, both straight and corrugated NWs had the thermal conductivity of 44 ± 1 Wm$^{-1}$K$^{-1}$. This value agrees with recent simulations[32] predicting about 47 Wm$^{-1}$K$^{-1}$ for both types of NWs with the same dimensions. However, this result contrasts with the measurements by Poborchii *et al.*[33], who measured about 70% lower thermal conductivity in the corrugated NWs at room temperature and attributed it to the phonon interference effects due to the periodicity of the constrictions. Our experiments show that such interference is unlikely to build up because the surface scattering at room temperature is mostly diffusive, albeit not entirely. Thus, the reduction observed by Poborchii *et al.*[33] is likely to be due to the high surface roughness of their corrugated structures.

Finally, we performed experiments and simulations to probe the directions of phonon transport in NWs. Both simulations and experiments agree that phonon directions are isotropic in straight NWs and only slightly oriented along the NW axis in long corrugated NWs. Thus, the ballisticity of phonon transport in NWs does not imply direct phonon flights from hot to cold. This conclusion seems to contradict our experimental results on straight and serpentine NWs, in which the turns of the NW block the direct phonon paths. To resolve this paradox, we analysed free phonon flights obtained using Monte Carlo simulation and found that phonon motion resembles the Lévy walk behaviour. The Lévy walk description has already been proposed to explain heat conduction in alloys[34,35], including alloyed NWs[18]. Here, we describe the Lévy walk induced by the confined geometry of NWs, in which phonons experience long flights parallel and short flights perpendicular to the NW axis. This behaviour explains the experimental results: the phonon transport cannot be strongly directional because any point of a NW has both long parallel flights and short perpendicular flights; still, turns of serpentine NWs block the long parallel flights and thus reduce the thermal conductivity.

To conclude, heat conduction in silicon nanostructures is more ballistic than it is common to assume[8,13–15]. Our experiments and simulations show that quasi-ballistic heat conduction occurs even in micron-long NWs up to room temperature and should be even more pronounced at the shorter length scales. Thus, we believe that the quasi-ballistic heat conduction combined with the ability to control its flow[27] may yet improve thermal management of silicon-based microelectronics.

## Methods

**Sample preparation**

The samples were created on a silicon-on-insulator wafer with a top monocrystalline undoped (100) silicon layer of 145 nm in thickness. First, using electron-beam assisted physical vapour deposition (Ulvac EX-300), we deposited square ($5 \times 5$ µm) aluminium transducers of 70 nm in thickness, shown in blue in Fig. 1a. Next, using electron-beam lithography followed by the inductively coupled plasma etching (Oxford Instruments Plasmalab System100 ICP), we etched the top silicon layer around the transducers leaving only the NWs to connect the central island with the surrounding wafer. The NWs were positioned along (100) and (010) crystallographic directions. Finally, to suspend the structure, we exposed the wafer to the vapour of diluted hydrofluoric acid, which removed the buried oxide layer under the central island and the NWs. Figure 1b shows one of the samples with the area of removed buried oxide visible around the structure.

**Monte Carlo simulations**

In the simulation, phonon wave packets are approximated by particles, hereafter called simply phonons. The algorithm generates $10^4$ phonons at the cold side of a NW with the wavelengths ($\lambda$) assigned according to the Plank distribution calculated within Debye approximation at 4 K. The group velocity and polarization are assigned according to the phonon dispersion in bulk. The phonons start moving in random directions in a three-dimensional model of a NW (Supplementary Fig. 7). At the boundaries, phonons are scattered either specularly (i.e. elastically) or diffusely (i.e. in a random direction according to Lambert cosine law). The specular scattering probability is determined by $p = exp\,(-16\,\pi^2\,\eta^2\,\cos^2\alpha/\lambda^2)$, where the r.m.s. surface roughness ($\eta$) is 0.3 nm for the top and bottom surfaces[27] and 2 nm for the side surfaces (1 nm from the TEM measurements and 1 nm to account for a possible oxide layer). Albeit negligible, the impurity and Umklapp scattering processes are implemented via relaxation time approximation. The simulation continues until all generated phonons leave the NW through the opposite side, where the algorithm records the angle $\theta$ between the NW axis and phonon direction projection in *x-y* plane. The algorithm also records the length of the projection of phonon flights between diffuse scattering events on the NW axis, thus obtaining the distribution of ballistic flights along the NWs. The simulations were conducted for the temperature of 4 K because the most interesting experimental results are obtained at this temperature, and the Debye approximation holds most accurately at low temperatures.

**Author contribution**

R.A. prepared the samples, conducted the measurements, analysed the experimental results, implemented Monte Carlo model, conducted the simulations, and prepared the article; S.G contributed to the TEM sample preparation and Monte Carlo implementation; S.G, S.V., and M.N contributed to the interpretation of the results and preparation of the article.

**Acknowledgements**

This work was supported by Kakenhi (15H05869, 15K13270, and 18K14078), PRESTO JST (JPMJPR15R4), and Postdoctoral Fellowship program of Japan Society for the Promotion of Science. We also acknowledge Ryoto Yanagisawa for assistance with the sample fabrication.

# References

1. Waldrop, M. M. The chips are down for Moore's law. *Nature* **530,** 144–147 (2016).

2. Mack, C. The Multiple Lives of Moore's Law. *IEEE Spectr.* **52,** 31–37 (2015).

3. Pop, E. Energy dissipation and transport in nanoscale devices. *Nano Res.* **3,** 147–169 (2010).

4. Siemens, M. E. *et al.* Quasi-ballistic thermal transport from nanoscale interfaces observed using ultrafast coherent soft X-ray beams. *Nat. Mater.* **9,** 26–30 (2010).

5. Lee, J., Lim, J. & Yang, P. Ballistic phonon transport in holey silicon. *Nano Lett.* **15,** 3273–3279 (2015).

6. Johnson, J. A. *et al.* Direct measurement of room-temperature nondiffusive thermal transport over micron distances in a silicon membrane. *Phys. Rev. Lett.* **110,** 025901 (2013).

7. Hsiao, T.-K. *et al.* Observation of room-temperature ballistic thermal conduction persisting over 8.3 μm in SiGe nanowires. *Nat. Nanotechnol.* **8,** 534–8 (2013).

8. Hsiao, T. K. *et al.* Micron-scale ballistic thermal conduction and suppressed thermal conductivity in heterogeneously interfaced nanowires. *Phys. Rev. B* **91,** 035406 (2015).

9. Zhang, Q. *et al.* Thermal transport in quasi-1D van der Waals crystal Ta2Pd3Se8 nanowires: Size and length dependence. *ACS Nano* **12,** 2634–2642 (2018).

10. Xu, X. *et al.* Length-dependent thermal conductivity in suspended single-layer graphene. *Nat. Commun.* **5,** 3689 (2014).

11. Bae, M.-H. H. *et al.* Ballistic to diffusive crossover of heat flow in graphene ribbons. *Nat. Commun.* **4,** 1734 (2013).

12. Huang, B.-W. W., Hsiao, T.-K. K., Lin, K.-H. H., Chiou, D.-W. W. & Chang, C.-W. W. Length-dependent thermal transport and ballistic thermal conduction. *AIP Adv.* **5,** 053202 (2015).

13. Maire, J., Anufriev, R. & Nomura, M. Ballistic thermal transport in silicon nanowires. *Sci. Rep.* **7,** 41794 (2017).

14. Hertzberg, J. B., Aksit, M., Otelaja, O. O., Stewart, D. A. & Robinson, R. D. Direct measurements of surface scattering in Si nanosheets using a microscale phonon spectrometer: Implications for casimir-limit predicted by ziman Theory. *Nano Lett.* **14,** 403–415 (2014).

15. Raja, S. N. *et al.* Length scale of diffusive phonon transport in suspended thin silicon nanowires. *Nano Lett.* **17,** 276–283 (2017).

16. Hippalgaonkar, K. *et al.* Fabrication of microdevices with integrated nanowires for investigating low-dimensional phonon transport. *Nano Lett.* **10,** 4341–4348 (2010).

17. Xie, G. *et al.* Phonon surface scattering controlled length dependence of thermal conductivity of silicon nanowires. *Phys. Chem. Chem. Phys.* **15,** 14647–14652 (2013).

18. Upadhyaya, M. & Aksamija, Z. Nondiffusive lattice thermal transport in Si-Ge alloy nanowires. *Phys. Rev. B* **94,** 174303 (2016).

19. Park, M. & Kim, Y. Lattice thermal conductivity of pristine Si nanowires : classical nonequilibrium molecular dynamics study. *Nanoscale Microscale Thermophys. Eng.* **21,** 278–286 (2017).

20. Yang, N., Zhang, G. & Li, B. Violation of Fourier's law and anomalous heat diffusion in silicon nanowires. *Nano Today* **5,** 85–90 (2010).

21. Regner, K. T., Freedman, J. P. & Malen, J. a. Advances in Studying Phonon Mean-Free-Path-Dependent Contributions to Thermal Conductivity. *Nanoscale Microscale Thermophys. Eng.* **19,** 183–205 (2015).

22. Regner, K. T. *et al.* Broadband phonon mean free path contributions to thermal conductivity measured using frequency domain thermoreflectance. *Nat. Commun.* **4,** 1640 (2013).

23. Hu, Y., Zeng, L., Minnich, A. J., Dresselhaus, M. S. & Chen, G. Spectral mapping of thermal conductivity through nanoscale ballistic transport. *Nat. Nanotechnol.* **10,** 701–706 (2015).

24. Zhang, H., Hua, C., Ding, D. & Minnich, A. J. Length dependent thermal conductivity measurements yield phonon mean free path spectra in nanostructures. *Sci. Rep.* **5,** 9121 (2015).

25. Heron, J. S., Bera, C., Fournier, T., Mingo, N. & Bourgeois, O. Blocking phonons via nanoscale geometrical design. *Phys. Rev. B* **82,** 155458 (2010).

26. Weng, Q. *et al.* Near-Field Radiative Nanothermal Imaging of Nonuniform Joule Heating in Narrow Metal Wires. *Nano Lett.* **18,** 4220−4225 (2018).

27. Anufriev, R., Ramiere, A., Maire, J. & Nomura, M. Heat guiding and focusing using ballistic phonon transport in phononic nanostructures. *Nat. Commun.* **8,** 15505 (2017).

28. Verdier, M., Anufriev, R., Ramiere, A., Termentzidis, K. & Lacroix, D. Thermal

conductivity of phononic membranes with aligned and staggered lattices of holes at room and low temperatures. *Phys. Rev. B* **95,** 205438 (2017).

29. Parrish, K. D., Abel, J. R., Jain, A., Malen, J. A. & McGaughey, A. J. H. Phonon-boundary scattering in nanoporous silicon films: Comparison of Monte Carlo techniques. *J. Appl. Phys.* **122,** 125101 (2017).

30. Park, W. *et al.* Phonon conduction in silicon nanobeam labyrinths. *Sci. Rep.* **7,** 6233 (2017).

31. Blanc, C., Rajabpour, A., Volz, S., Fournier, T. & Bourgeois, O. Phonon heat conduction in corrugated silicon nanowires below the Casimir limit. *Appl. Phys. Lett.* **103,** 043109 (2013).

32. Verdier, M., Lacroix, D. & Termentzidis, K. Heat transport in phononic-like membranes: Modeling and comparison with modulated nano-wires. *Int. J. Heat Mass Transf.* **114,** 550–558 (2017).

33. Poborchii, V., Morita, Y., Hattori, J., Tada, T. & Geshev, P. I. Corrugated Si nanowires with reduced thermal conductivity for wide-temperature-range thermoelectricity. *J. Appl. Phys.* **120,** 154304 (2016).

34. Vermeersch, B., Carrete, J., Mingo, N. & Shakouri, A. Superdiffusive heat conduction in semiconductor alloys. I. Theoretical foundations. *Phys. Rev. B* **91,** 085202 (2015).

35. Vermeersch, B., Mohammed, A. M. S., Pernot, G., Koh, Y. R. & Shakouri, A. Superdiffusive heat conduction in semiconductor alloys. II. Truncated Lévy formalism for experimental analysis. *Phys. Rev. B* **91,** 085203 (2015).